\documentclass[conference]{IEEEtran}
\IEEEoverridecommandlockouts
\usepackage{cite}
\usepackage{amsmath,amssymb,amsfonts}
\usepackage{graphicx}
\usepackage{subfig}
\usepackage{textcomp}
\usepackage{xcolor}
\usepackage{amsmath}
\usepackage[ruled,linesnumbered]{algorithm2e}
\usepackage{array}
\usepackage{setspace}
\usepackage[normalem]{ulem}
\def\BibTeX{{\rm B\kern-.05em{\sc i\kern-.025em b}\kern-.08em
    T\kern-.1667em\lower.7ex\hbox{E}\kern-.125emX}}

\begin{document}
\SetKwComment{Comment}{/* }{ */}

\title{Network-aware Prefetching Method for Short-Form Video Streaming}

\author{
\IEEEauthorblockN{Duc Nguyen}
\IEEEauthorblockA{
\textit{Tohoku Institute of Technology}\\
Sendai, Japan \\
ducnguyen@tohtech.ac.jp}
\and
\IEEEauthorblockN{Phong Nguyen, Vu Long, Truong Thu Huong}
\IEEEauthorblockA{
\textit{Hanoi University of Science and Technology}\\
Hanoi, Vietnam \\
huong.truongthu@hust.edu.vn
}
\and
\IEEEauthorblockN{Pham Ngoc Nam}
\IEEEauthorblockA{
\textit{VinUniversity}\\
Hanoi, Vietnam\\
nam.pn@vinuni.edu.vn
}

}

\maketitle

\begin{abstract}
Recent years have witnessed the rising of short-form video platforms such as TikTok. Apart from conventional videos, short-form videos are much shorter and users frequently change the content to watch. Thus, it is crucial to have an effective streaming method for this new type of video. In this paper, we propose a resource-efficient prefetching method for short-form video streaming. Taking into account network throughput conditions and user viewing behaviors, the proposed method dynamically adapts the amount of prefetched video data. Experiment results show that our method can reduce the data waste by $37\sim52\%$ compared to other existing methods.
\end{abstract}


\section{Introduction}
Video streaming is one of the most important mobile applications nowadays. According to \cite{Ericsson2021}, video traffic accounts for 69\% of the total mobile network data traffic as of 2021, and is predicted to increase to 79\% by 2027. In recent years, we have witnessed the rising of short-form videos, whose duration is typically a few minutes or shorter. Short-form video platforms such as TikTok is attracting millions of users sharing various types of user-generated short-form videos~\cite{Tiktok}.

Short-form videos are different from conventional videos in many aspects. Especially, viewers have a limited control over not only video playback but also video content. Based on various input data such as the user's past viewing behaviors, short-form streaming platforms continuously recommend videos to users~\cite{TikTok_recomendation}. The user agent then downloads the recommended videos from a streaming server. In short-form video streaming, the videos are always displayed in a full-screen mode, and the playback controls such as pause, resume are typically not available. To stop watching the current video, users need to scroll to the next or previous video. In the event of scrolling, previously downloaded data of the current video is discarded~\cite{Zhang2022}.  

So far, just a few efforts have been made to optimize short-form video streaming over resource-constrained networks. Recent measurement studies found that commercial short-form video platforms employ a simple streaming approach where videos are sequentially downloaded to the user device. Unfortunately, this approach causes a significant waste in the network resource~\cite{Zhang2022}. To tackle this problem, buffer-based streaming approaches have been proposed~\cite{Zhang2021,Zhang2022b,He2020}. The main idea is to constraint the amount of prefetched video data to reduce data wastage when users scroll to the next video. The previous studies employ deep learning approaches to learn the optimal value of the required buffer size. The primary problem with these approaches is that a huge amount of user data (i.e., user scrolling behaviors) must be collected in order to train the deep learning model. In practice, collecting user data is a non-trivial task when there are more and more user data protection laws such as GDPR~\cite{GDPR} are being applied.

In this paper, we propose a novel resource-efficient prefetching scheme for short-form video streaming over mobile networks. The proposed method has the following key features:
\begin{itemize}
    \item First, our method is simple and requires only local user data. The buffer size is dynamically adapted according to network conditions to jointly minimize data wastage and re-buffering times. 
    \item Second, parallel prefetching is employed to prefetch video segments of not only the current video but also next videos in the playlist. This can significantly reduce the start-up delay when users scroll videos.
\end{itemize}
Trace-driven evaluation shows that the proposed method can significantly reduce the data wastage, the re-buffering time and start-up delay. The remaining of the paper is structured as follows. A summary of related studies is presented in Section~\ref{sec:relatedwork}. An overview of short-form video streaming problem is given in Section~\ref{sec:problem}. The proposed method is given in Section~\ref{sec:proposedmethod}, followed by an evaluation in Section~\ref{sec:evaluation}. Finally, the paper is concluded in Section~\ref{sec:conclusion}.

\section{Related Work}\label{sec:relatedwork}
Recent measurements on commercial short-form video sharing platforms have found that a significant amount of video data is wasted in practice~\cite{He2020,Zhang2022}. This is mainly because of the simple video download strategy adopted by commercial platforms. In particular, videos are downloaded in sequence and the download of the next video begins only after the download of the current video has been complete. To address this problem, existing works have proposed machine learning-based methods ~\cite{He2020, Guo2021,Zhang2021,Zhang2020,Zhang2022b}. 

\textcolor{black}{In~\cite{Zhang2020}, the authors proposed an adaptive prefetching approach (APL) to minimize the waste and the stall time of short-form video streaming. At each consecutive time slot, the video downloader chooses which video to download for an amount of time. The result is evaluated based on predicting users' watching time of the past videos. The authors used a sliding window to predict the viewing duration according to the maximum of 5 videos in the past. The problem of this solution is that the sliding window has too few arguments to predict thus giving a unconvincing predicting results and led to poor implementation of different users behaviors. }\textcolor{black}{The LiveClip proposed in~\cite{He2020}  deals with the existing problem of the APL method~\cite{Zhang2020}. The authors eliminate the old sliding window method and replace it with the reinforcement learning algorithm to predict user viewing duration. This approach thus gives a much more precise result in the expected watching duration. Since the users' short-term viewing is highly predictable, the waste time is reduced significantly in comparison with the APL method.}

In~\cite{Zhang2021}, the authors propose a wastage-aware short-form video streaming method (WAS). In the WAS method, the download of the next segment is scheduled so that the buffer occupancy at any given time is always lower than a threshold. This can help reduce the data wastage in case the user switches from one video to another one during the playback process. In addition, borrowing  the idea of Bitrate Adaptive Streaming, WAS dynamically adjusts the video bitrate based on the available throughput. For that, multiple versions with different bitrates of a video are prepared and stored in the server in advance. The main problem with these methods is that they require a large amount of user data. In practice, collecting user data is a non-trivial task due to many restrictions. In~\cite{Zhang2022b}, the authors proposes DUASVS, a deep learning-based short-form video streaming method.  In DUASVC, once the prefetch duration of the current video reaches a threshold, the video player will shift to fetching the next video. An actor-critic network is trained to decide the threshold. As a result, this method also requires collecting a lot of training data. \textcolor{black}{Another problem of these above methods is that they only take into consideration the expected viewing duration. However, a key factor that affects the users experience is the start-up delay and the re-buffering time or the stall time which causes by the fluctuation of the network, thus the future bandwidth prediction should also be focused on.}

\begin{figure}
    \centering
    \subfloat[System Architecture]{
    \includegraphics[width=\columnwidth]{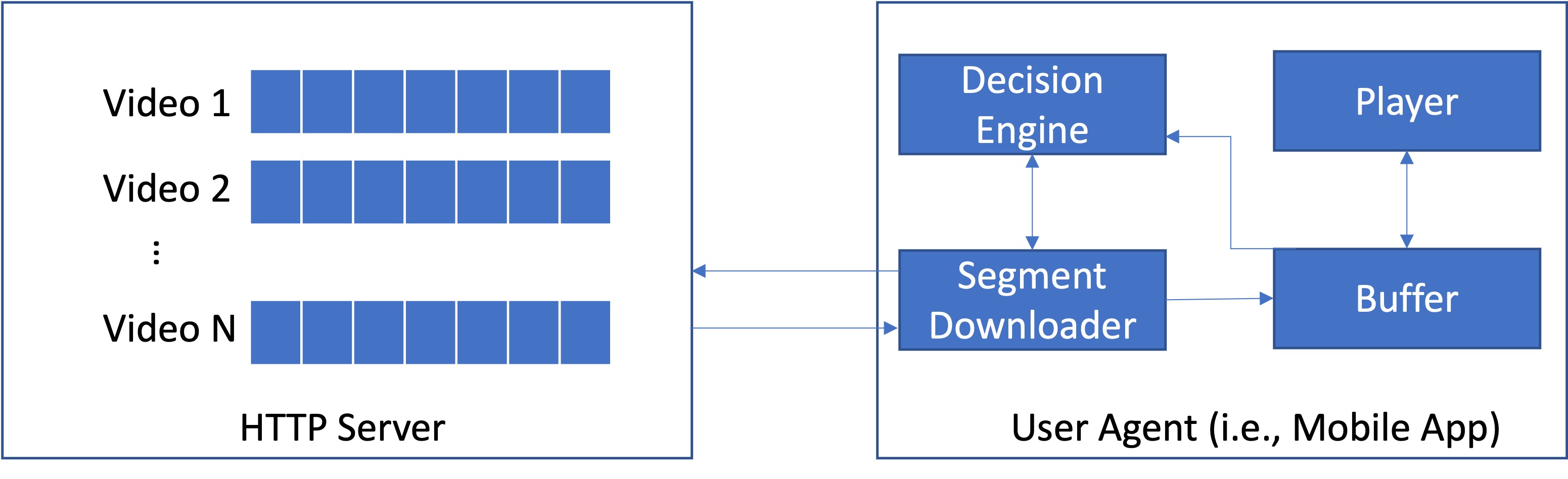}
    }\hfill
     \subfloat[User Session]{
    \includegraphics[width=\columnwidth]{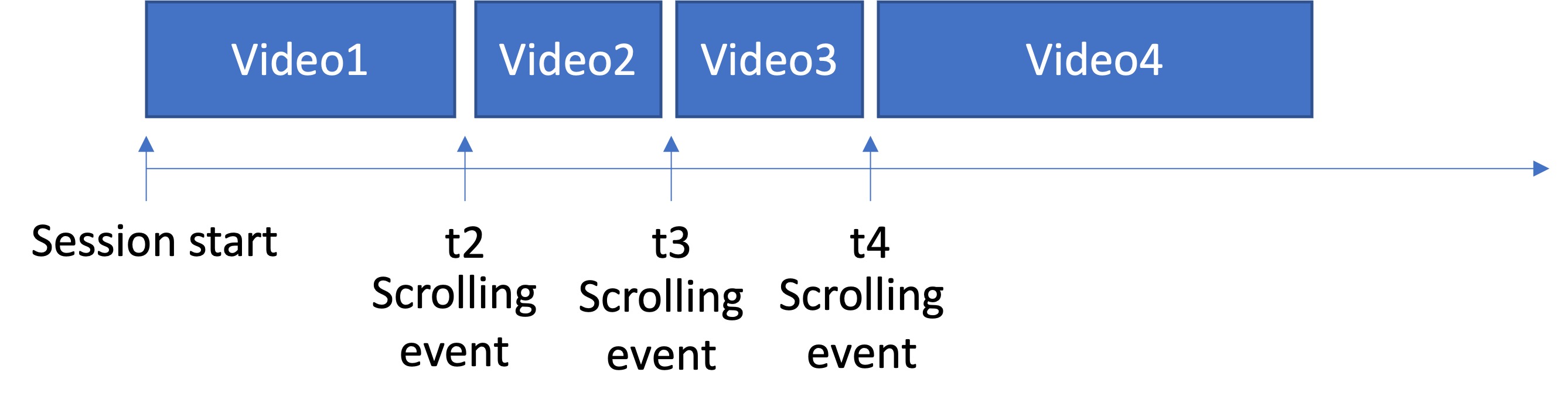}
    }
    \caption{Short-Form Video Streaming System}
    \label{fig:system_arch}
\end{figure}

\section{Overview of Short-Form Video Streaming}\label{sec:problem}
The general architecture of a short-form video streaming system is shown in Fig.~\ref{fig:system_arch}a. We consider the scenario in which a user is watching videos on short-video platforms on his/her mobile device (e.g., smartphones) over wireless networks (e.g., Wifi, 4G/5G). The videos are stored in a HTTP server where each video is encoded and temporally divided into small chunks called \textit{segments} with a same playback duration. To download a video from the server, the segment downloader of the user equipment (UE) sends HTTP requests for individual segments of the video to the HTTP server. Upon receiving the requests, the HTTP server responses with the requested segments to the client via HTTP response messages. The decision on which segment should be downloaded is made by a decision engine. 

After sending the request, the downloader receives the requested video segment and stores it in the buffer. The player then decodes the arrived video segments, and displays them on the UE's screen. A key characteristic of short-form video streaming is that users frequently change the content to watch. In general, users will spend more time on videos of his/her interest. Especially, users can only scroll to the previous or next videos. A typical user session is shown in Fig.~\ref{fig:system_arch}b. 

In fact, there are 3 factors affecting Quality of Experience (QoE) in short-form video streaming. The first factor is video quality. Because video data is transmitted over HTTP protocol, quality degradation caused by network packet losses can be completely avoided. Thus, video quality is dependent on the encoding process. Variable Bitrate Encoding (VBR) ensures stable quality across video segments, whereas Constant Bitrate Encoding (CBR) yields varying video qualities. The second factor is re-bufferings in which video playback is temporarily stopped due to lack of video data in the buffer. This problem happens when a video segment is available at the user agent later than its playback deadline. The main cause of this problem is sudden drops under an available network throughput condition. Re-buffering events have a significantly negative impact on user-perceived quality of video streaming~\cite{HuyenQoE2016}. The third key factor is start-up delay, \textcolor{black}{which is the time when a user scrolls at a video until the playback of the video begins}. Since users frequently scroll videos, low start-up delay is of especially important to ensure user satisfactory ~\cite{Zhang2022}. If one or more segments of a video have been downloaded already, then the video can be started instantly without any start-up delay. Otherwise, users must wait at least until the first segment of the considered video has been completely downloaded. 

In addition to providing high Quality of Experience to users, short-video streaming systems must also reduce the data wastage as much as possible. In particular, when users scroll to previous or next videos, all data of the current video that already been downloaded will be discarded. Recent measurements~\cite{He2020,Zhang2022} on the commercial short-form video streaming platforms found that nearly 45\% of the downloaded video data is eventually discarded. Such a high ratio of data wastage is not desirable for not only users but also service providers.

\section{Proposed Method}\label{sec:proposedmethod}
\subsection{Problem Formulation}
We define a viewing session as the time from when a user opens the short-form streaming app to when \textcolor{black}{he/she} closes the app. During the viewing session, the user watches a set of $N$ videos $\{v_1, v_2, \dots, v_N\}$ in sequential order. Video $v_i$ has a bitrate of $R_i$ and a playback duration of $L_i$. Video $v_i$ is divided into $M_i=\frac{L_i}{\tau}$ segments with $\tau$ is the playback duration of a segment. Let $t_i~(1\leq i \leq N)$ denote the time the user spends watching video $v_i$. We also suppose that the user only scrolls to the next video in the playlist. The playback of a video can start when the number of buffered segments of the video reaches an initial threshold $B_0$.
\begin{algorithm}[t]
\caption{Network-aware Segment Prefetching}\label{alg:proposed}

$i^{cur} \gets 1$\ \Comment*[r]{Current video id}
\While{$t < T^{sess}$}{
  Update value of $B_1$ and $K$ using Eq.~(\ref{eq:calculate_B1_a})(\ref{eq:calculate_B1_b})(\ref{eq:calculate_K})\;
  \eIf{$B(i^{cur}, t^{now})< B_1$}{
    Prefetch the next segment of current video\;
    Update buffer size of current video using Eq. (1)\;
  }{
    \For{$k\gets0$ \KwTo $K$}{
    \If{$i^{cur} + k \leq N$ and $B(i^{cur}+k, t^{now}) < B_1$}{
    Prefetch the next segment of video $(i^{cur} + k)$\;
    Update buffer size of video $(i^{cur} + k)$ using Eq. (1)\;
    \textbf{break}\;
    }
    }
  }
}
\end{algorithm}

Let $D^t=\{v^t, k^t\}$ denote the download decision made at time $t$. The buffer size $B(v^t, k^t)$ of video $v^t$ after downloading segment $k^t$ is given by,
\begin{equation}
    B(v^t, k^t) = \max(B(v^t, k^t-1) - \frac{\tau R_{v^t}}{Thrp(t)}, 0) + \tau,
\end{equation}
where $Thrp(t)$ denotes the network throughput at time $t$. If the segment download time is higher than the current buffer size, then a re-buffering event will occur. The re-buffering time at segment $k^t$ of video $v^t$ is thus given by:

\begin{equation}
I(v^t, k^t) =  \max(B(v^t, k^t-1) - \frac{\tau R_{v^t}}{Thrp(t)}, 0).
\end{equation}
When the user scrolls to the next video, if there are $B_0$ segments of that video in the buffer, then the playback of the next video can start immediately. Otherwise, the user agent must wait until the first $B_0$ segments are completely downloaded. This delay time is referred to as \textit{start-up delay}. The start-up delay $D(i)$ of video $v_i$ can be computed as follows.
\begin{equation}
    D(i) = \max(t^a(i, B_0) - t^s(i), 0)
\end{equation}
where $t^s(i)$ denotes the time the user scroll to video $v_i$ and $t^a(v_i, B_0)$ denotes the time where the first $B_0$ segments of video $v_i$ is \textcolor{black}{completely} downloaded. In addition, all the segments of the current video will be discarded when the user scrolls to the next video. The amount of discarded data $W(i)$ of video $i$ is equal to the buffer size at scrolling time $t^s(s+1)$.
\begin{equation}
    W(i) = B(v_i, t^s(i+1))
\end{equation}
To this end, the short-form video streaming problem can be stated as follows.

\textit{Under varying network conditions and dynamic user behaviors, decide the prefetch schedule $\{D(t)\}$ to download videos in the current playlist so as to maximize the overall quality $OQ$ which is a function of data wastage, video bitrates, re-bufferings and start-up delays}
\begin{equation}
\begin{split}
    OQ = f(\{R(i), 1\leq i \leq N\}, \\ 
    \{I(i), 1\leq i \leq N\}, \\ \{D(i), 1\leq i \leq N\}\\
    \{W(i), 1\leq i \leq N\}
    )
\end{split}
\end{equation}
Because both interruptions and start-up delay have negative impacts on user QoE, they should be as small as possible in order to providing high Quality of Experience.

\subsection{Network-aware Prefetching Method}
In this part, we present a novel prefetching method for short-form video streaming to optimize both user's Quality of Experience and network resources. The proposed method is summarized in Algorithm~\ref{alg:proposed}. To reduce the data wastage, the proposed method prefetches segments of the current video so that the amount of buffered video data at any time is approximately $B_1$ seconds. The buffered video data can also help mitigating re-bufferings under network throughput reductions. It can be noted that the smaller the value of $B_1$ is, the lower the amount of wastage would become. However, choosing too small value of $B_1$ might result in re-bufferings under significant drops under a network throughput condition. Thus, we propose to dynamically adjust the number of prefetched segments based on recent network conditions.  First, the average network throughput over the last $T$s is computed.
\begin{equation}\label{eq:calculate_B1_a}
    \alpha = \frac{1}{T} \sum_{t=t_{now}-T}^T Thrp(t)
\end{equation}
Here, $Thrp(t)$ denotes the network throughput sample measured at time $t$. In our system, a throughput sample is calculated after a video segment is completed downloaded by dividing the segment size to the download time. The number of prefetched segments $B_1$ is then computed as follows.
\begin{equation}\label{eq:calculate_B1_b}
    B_1 = \begin{cases}
  4,  & \alpha \leq 1.5R_i\\
  3,  & 1.5R_i < \alpha \leq 2.5R_i\\
  2,  & \textbf{otherwise}
\end{cases}
\end{equation}
\begin{figure}[t]
    \centering
    \subfloat[Trace \#1]{
    \includegraphics[width=0.33\columnwidth]{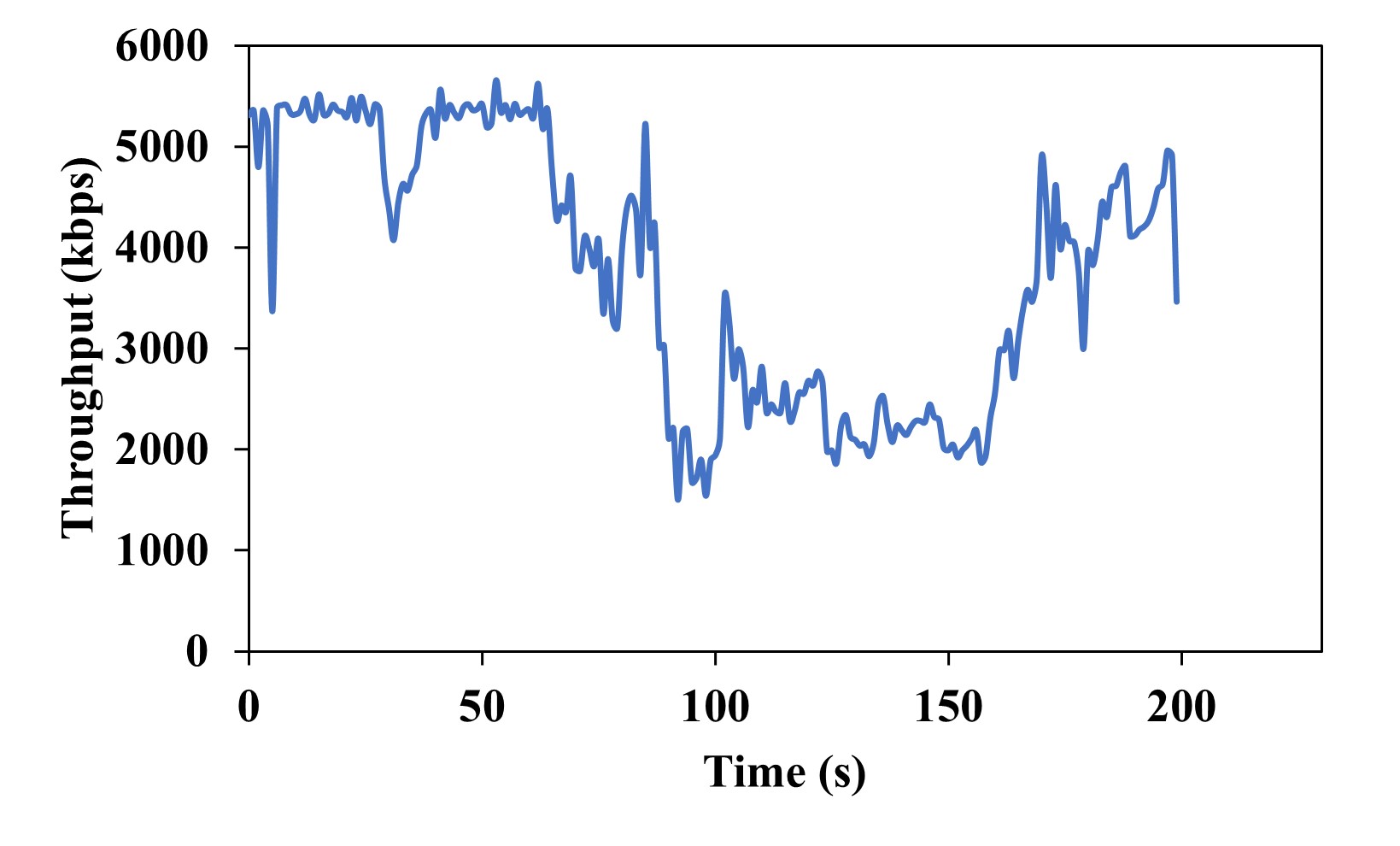}
    }
    \subfloat[Trace \#2]{
    \includegraphics[width=0.33\columnwidth]{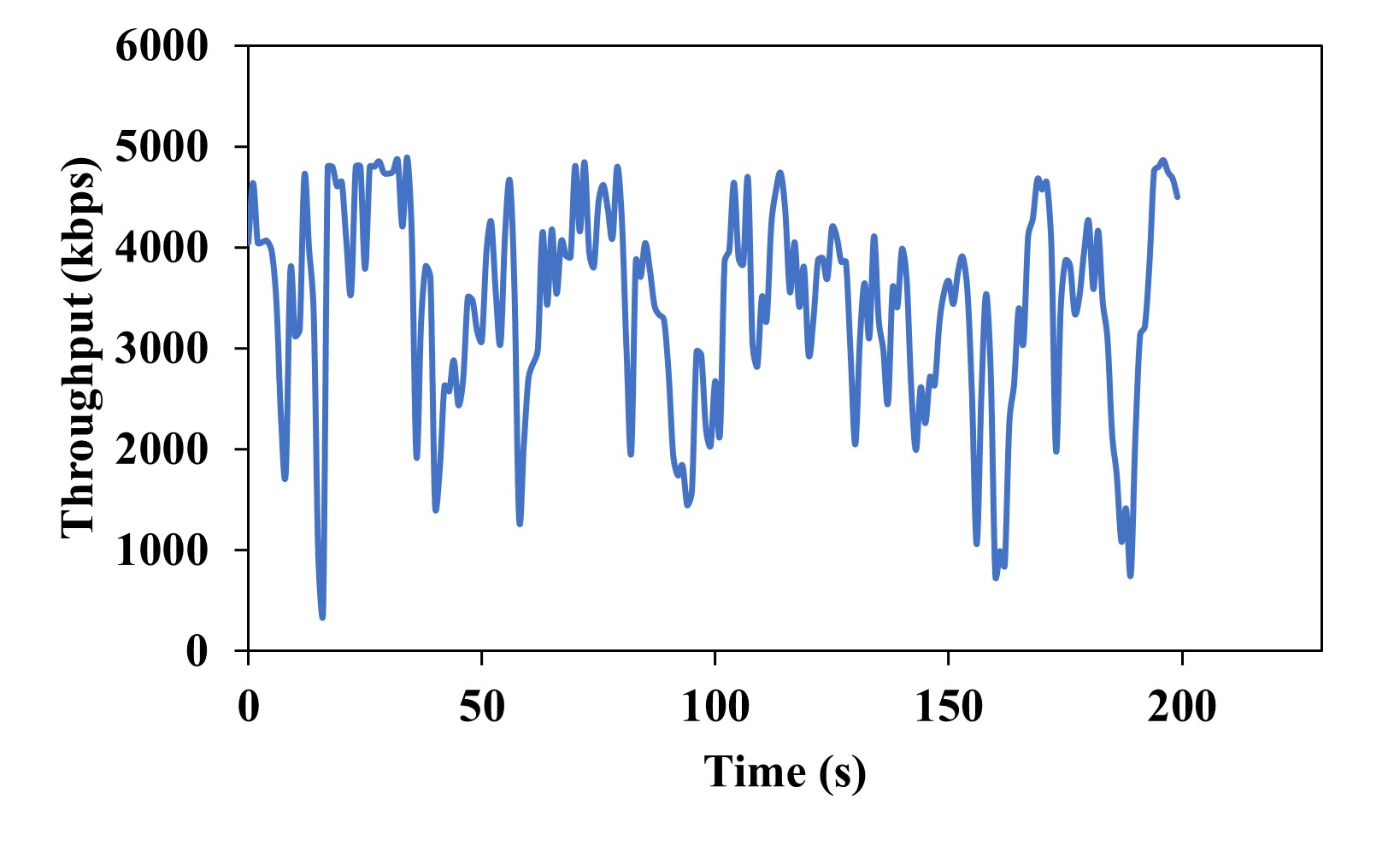}
    }
    \subfloat[Trace \#3]{
    \includegraphics[width=0.33\columnwidth]{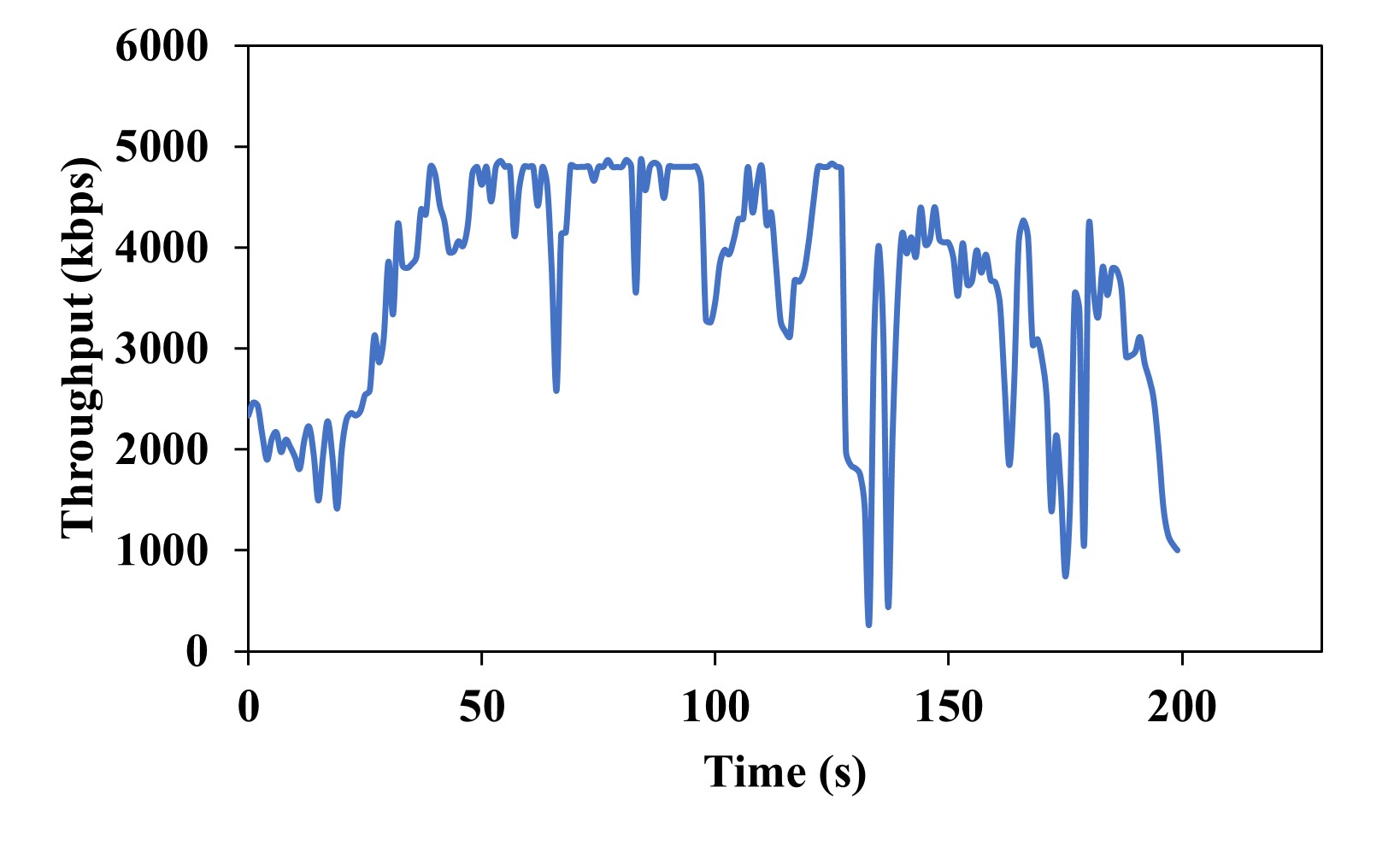}
    }
    \caption{Network throughput traces.}
    \label{fig:bw_trace}
\end{figure}
\begin{figure*}[t]
    \centering
    \subfloat[Waste (s)]{
    \includegraphics[width=0.3\textwidth]{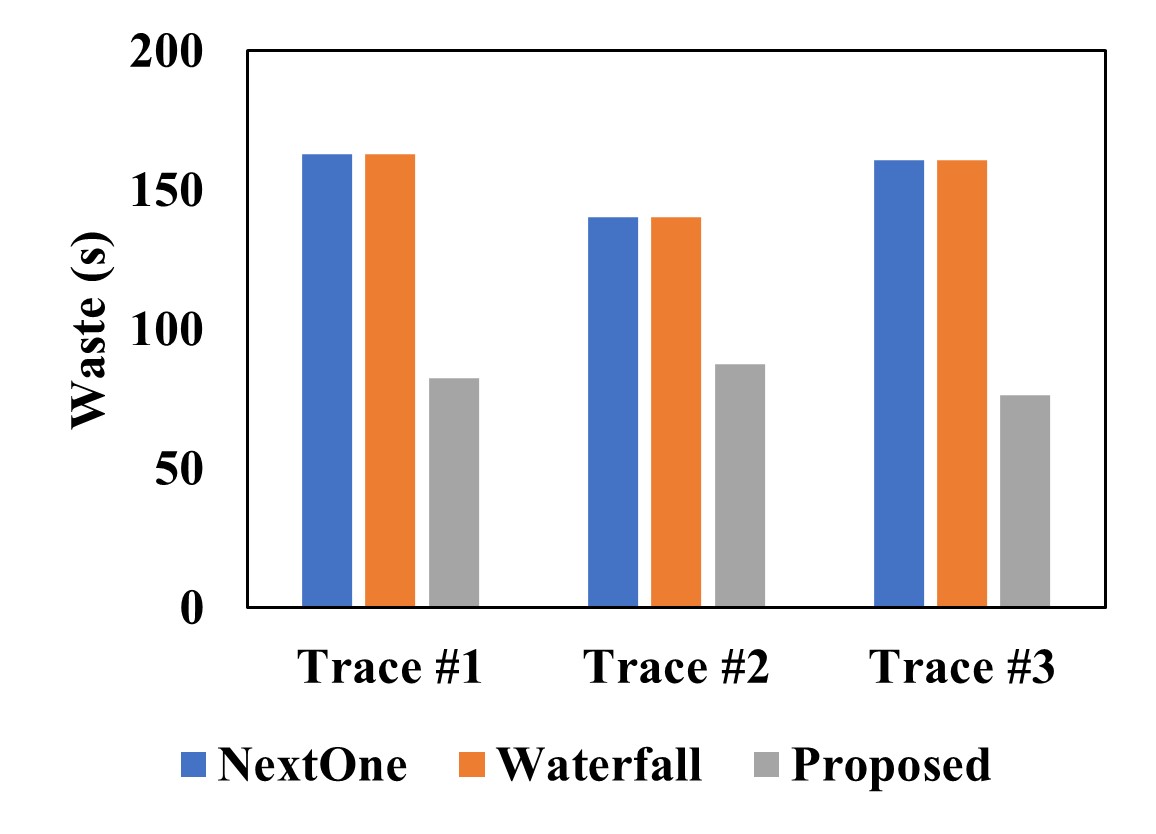}
    }
    \hfil
    \subfloat[Start-up delay (s)]{
    \includegraphics[width=0.3\textwidth]{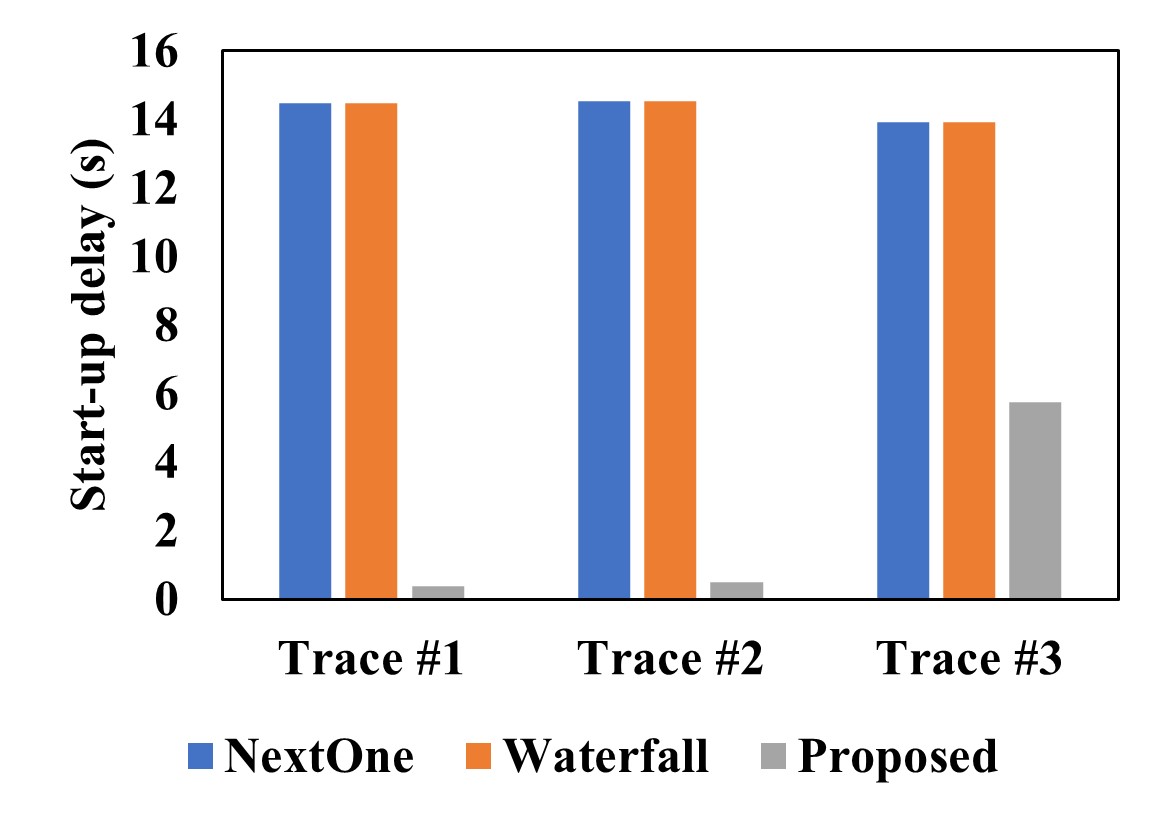}
    }
    \hfil
    \subfloat[Re-buffering time (s)]{
    \includegraphics[width=0.3\textwidth]{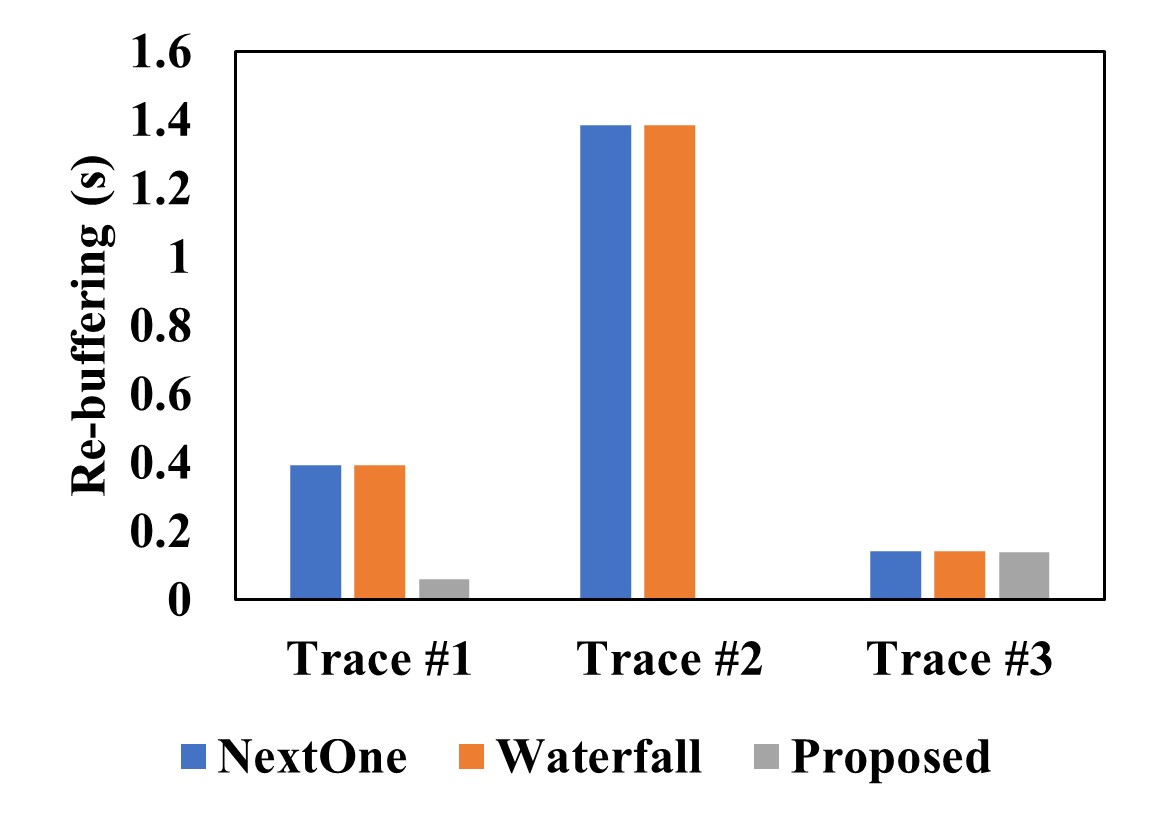}
    }
    \caption{Performance of the proposed method and reference methods under user trace \#1.}
    \label{Result_user_trace_1}
\end{figure*}
\begin{figure*}[t]
    \centering
    \subfloat[Waste (s)]{
    \includegraphics[width=0.3\textwidth]{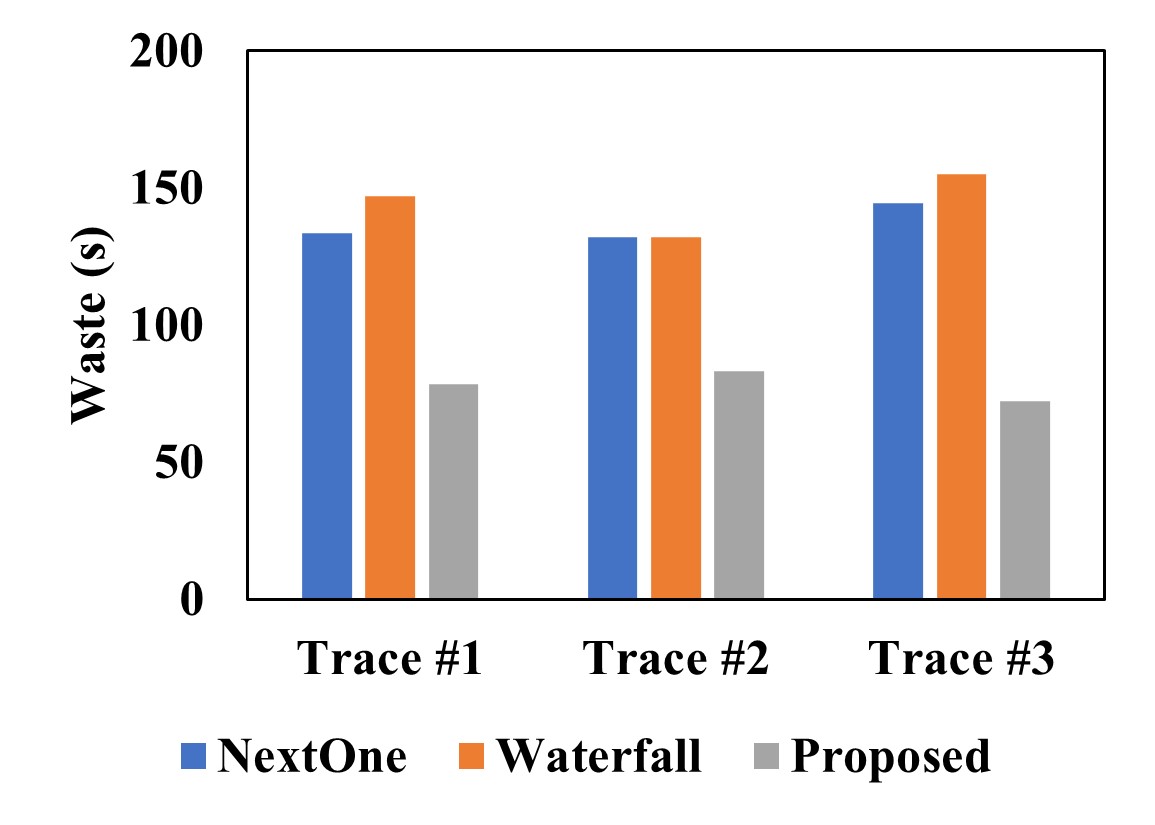}
    }
    \hfil
    \subfloat[Start-up delay (s)]{
    \includegraphics[width=0.3\textwidth]{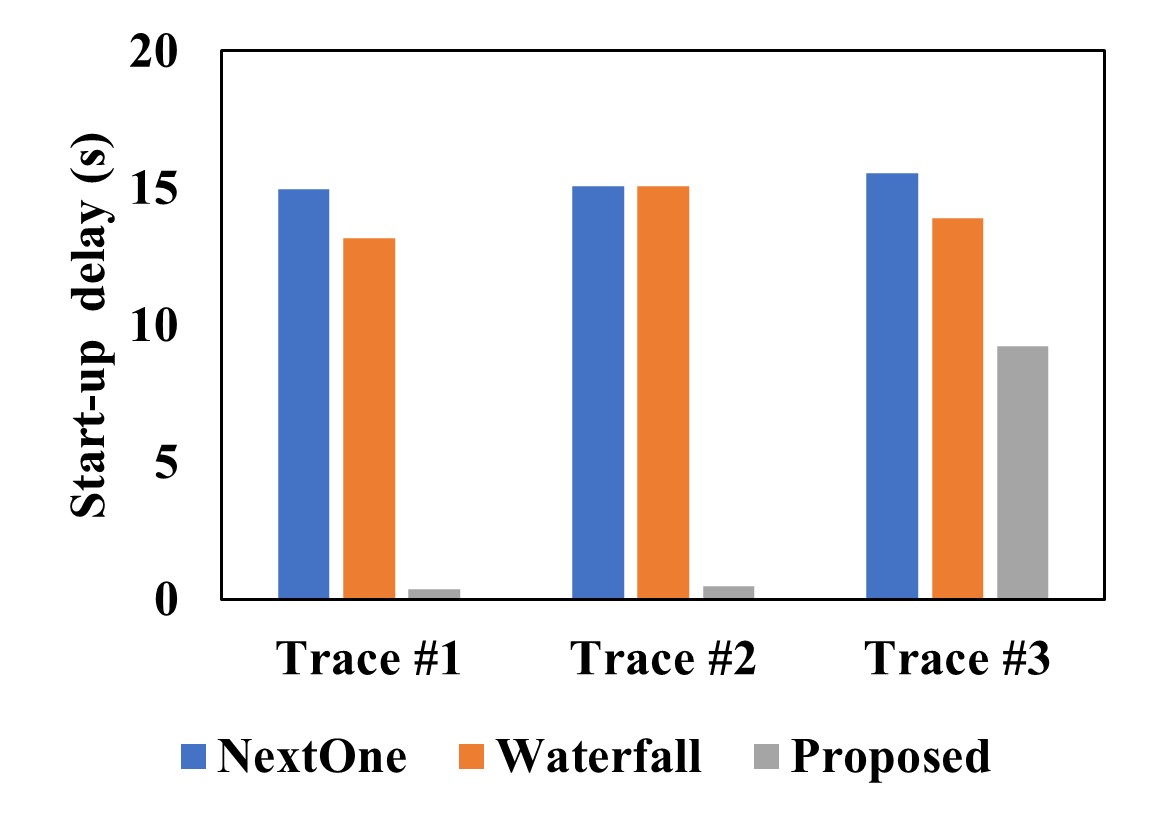}
    }
    \hfil
    \subfloat[Re-buffering time (s)]{
    \includegraphics[width=0.3\textwidth]{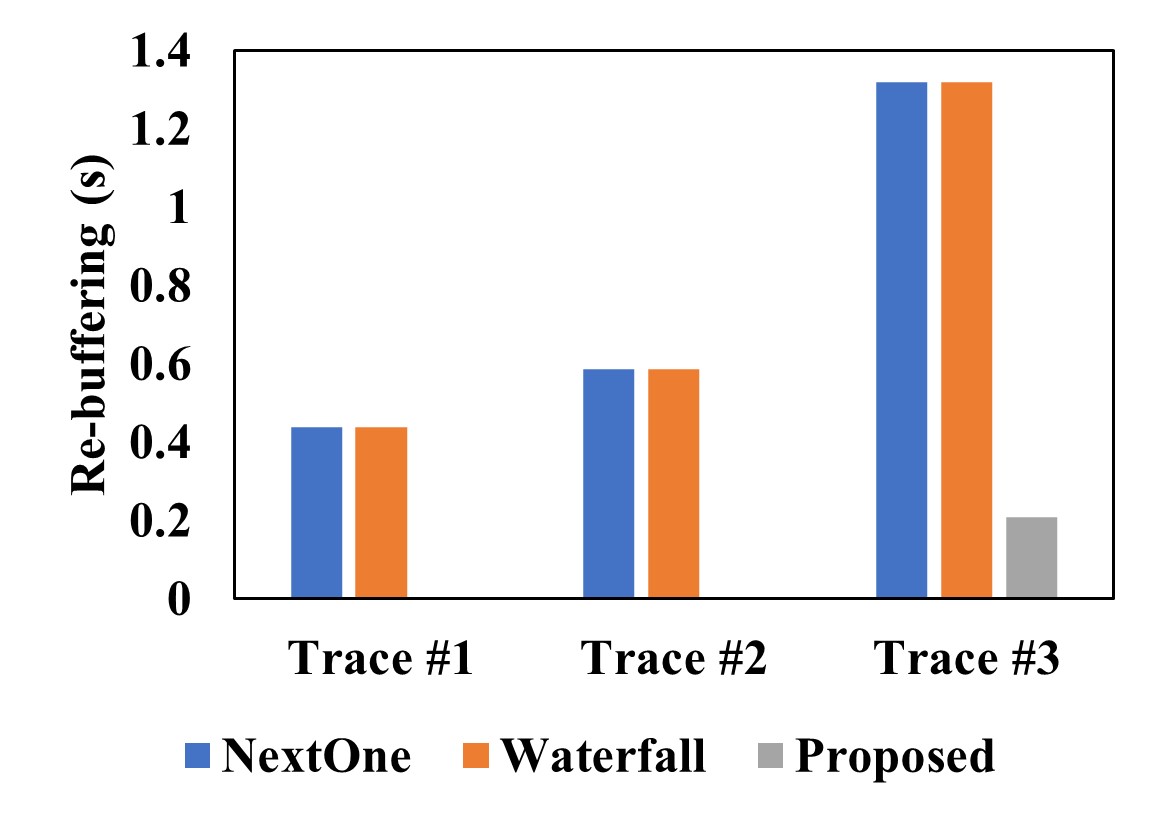}
    }
    \caption{Performance of the proposed method and reference methods under user trace \#2.}
    \label{Result_user_trace_2}
\end{figure*}
It can be noted that, the higher the value of $\alpha$ is, the lower the value of $B_1$ would become with the minimum buffer size is two segments. During the streaming session, our algorithm continuously compares the buffer size of the current video to $B_1$. If the current buffer size is smaller than $B_1$, then our method will download next segments of the current video until the buffer size exceeds the threshold $B_1$. Otherwise, our method will prefetch a segment of a next video in the current playlist. This is to facilitate zero start-up delay when users scroll videos. Specifically, the algorithm considers $K$ videos next to the current video in the playlist. The next segment of the first video of which the buffer size is still lower than threshold $B_1$ will be fetched. Similar to the buffer size $B_1$, the value of $K$ is also decided using the average network throughput $\alpha$ as follows.
\begin{equation}\label{eq:calculate_K}
    K = \begin{cases}
  7,  & \alpha \leq 1.5R_i\\    
  4,  & 1.5R_i < \alpha \leq 2R_i\\
  7,  & 2R_i < \alpha \leq 2.5R_i\\
  12,  & \textbf{otherwise}
\end{cases}
\end{equation}

\section{Performance evaluation}\label{sec:evaluation}
\subsection{Experimental Settings}
In our experiments, we take into consideration videos that have a length of 15 seconds with a constant bitrate of 2Mbps (so that each video has 30Mb in size). The videos are then individually divided into 1-second-long segments. The initial buffer threshold $B_0$ is 1 segment. We use three network bandwidth traces from mobile networks to simulate real network conditions as shown in Fig.~\ref{fig:bw_trace}. \textcolor{black}{Each bandwidth trace contains an array of network-throughput per seconds for 200 seconds. The first network trace simulates a relatively stable network with average throughput of 4500Kbps. The second and the third traces show highly fluctuated networks with average throughput of 2000Kbps.} For the user scrolling behaviors, \textcolor{black}{ since collecting users data is neither easy nor appropriate due to many restrictions, we simulate these by generating two user behavior traces with different properties. These two user traces are generated following Gaussian distribution. The user behavior trace consists of an array of numbers, each represent the watching time of the corresponding video in a sequential order}. User trace \#1 represents type of user who would likely to skip videos moderately (the mean and standard deviation of Gaussian distribution are 12 and 6, respectively). For user trace \#2, it represents people who scroll videos frequently (in this case, the mean value is 6 and the standard deviation value is 3). \textcolor{black}{Both user traces have total time of approximately 3 minutes.}

The proposed method is then compared with two reference methods \textbf{NextOne} and \textbf{Waterfall} that can be described as follows.
\begin{itemize}
   \item \textbf{NextOne}: In this method, all segments of the currently viewing video are buffered. The next video will NOT be buffered until the current video is completely downloaded. The number of next videos that will be automatically buffered and stored in the cache is limited to one.
    \item \textbf{Waterfall}: Similar to the \textbf{NextOne} method, the next videos will only be downloaded if the current video is finished downloading. The key difference here is that the number of next videos that can be buffered is increased to two.
\end{itemize}

\subsection{Experimental Results}
In our experiment, we measure 3 performance metrics, as follows:
\begin{itemize}
\item \textbf{Waste time}: the video time buffered that the user never watched by skipping to the next video. 
\item \textbf{Start-up delay}: The amount of time from the moment the user switched to the next video until it is playable. 
\item \textbf{Re-buffering time}: the time that the user has to wait until the video playback starts.
\end{itemize}

Figure \ref{Result_user_trace_1} and Figure \ref{Result_user_trace_2} show the overall outcome of waste time, start-up delay and re-buffering time of our proposed method in comparison with the other two reference methods. The results can be seen varied based on 3 different throughput traces. We can observe from the first throughput trace ( i.e., throughput trace \#1 ) that it has a relatively high throughput and fluctuates slightly. Therefore, our method clearly outperforms both reference methods in terms of all three metrics. The start-up delay and re-buffering time of our proposed method is much lower since it reduces up to more than 90\% compared to both \textbf{NextOne} and \textbf{Waterfall}. And the waste time is decreased by $41\sim49\%$. 

In the second case (i.e., throughput trace \#2), the network throughput fluctuates remarkably, the approximate average speed is more than 2000Kbps. We can conclude from Figure \ref{Result_user_trace_1} and \ref{Result_user_trace_2} that the waste time is reduced by 37\% in comparison with \textbf{Waterfall} and \textbf{NextOne}. In this second case, our method's result in start-up delay and re-buffering time out-performs the other two reference methods by 96\%.

Finally, throughput trace \#3 represents a moderately fluctuating network bandwidth. Because its network throughput in the first 25 seconds is quite low (less than 2000Kbps), the start-up delay is inevitable. However, our method can still reduce this value by $33\sim58\%$ in both user trace scenarios (i.e., user trace 1 and 2). Furthermore, our algorithm is still performing reasonably good with the remaining 2 metrics. The waste time of our proposed method in user trace \#1 is reduced by 52\%,  and in user trace \#2 is reduced by $50\sim53$\%. The re-buffering time in user trace \#1 is quite similar in all methods and in user trace \#2 with faster scrolling speed of user behavior, the re-buffering time is improved significantly, it is less than 84\% compares to the two reference methods which are \textbf{Waterfall} and \textbf{NextOne}. 

\section{Conclusions}\label{sec:conclusion}
In  this  paper,  we  have demonstrated  a  method for streaming short-form videos to users over time-varying networks. Our proposed method dynamically adjusts the number of buffered segments according to recent network conditions. Also, the proposed method prefetches not only segments of the current video, but also those of next videos in the playlist. Experimental results show that our proposed method can significantly reduce the data wastage, re-buffering time, and start-up delay compared to two reference methods. In future work, we will develop method for multiple users in a same network.
\section*{Acknowledgment}
This work was funded by Vingroup and supported by Vingroup Innovation Foundation (VINIF) under project code VINIF.2020.DA03 and Competitive Fund from Tohoku Institute of Technology, Japan. 
\bibliographystyle{./bio/IEEEtran}
\bibliography{./bio/ref.bib}
\end{document}